\documentclass[sigconf,natbib=false]{acmart}

\setcopyright{none}
\renewcommand\footnotetextcopyrightpermission[1]{}
\AtBeginDocument{%
  }

\usepackage{comment}
\usepackage{booktabs} % For formal tables
\usepackage{pifont}
\usepackage{array}

\usepackage{tikz}
\usetikzlibrary{plotmarks}
\usepackage{graphicx}
\usepackage{amsmath}
\usepackage{pgfplots}

\RequirePackage[
  datamodel=acmdatamodel,
  style=acmnumeric,
  ]{biblatex}

\begin{document}
\title{Large Language Models for Analyzing Enterprise Architecture Debt in Unstructured Documentation}
  
\renewcommand{\shorttitle}{Large Language Models for Analyzing Enterprise Architecture Debt}

\author{Christin Pagels}
\affiliation{%
  \institution{Stockholm University}
  \city{Stockholm} 
  \country{Sweden} 
}
\email{chpa1850@student.su.se}

\author{Simon Hacks}
\orcid{0000-0003-0478-9347}
\affiliation{%
  \institution{Stockholm University}
  \city{Stockholm} 
  \country{Sweden}  
}
\email{simon.hacks@dsv.su.se}

\author{Rob Henk Bemthuis}
\orcid{0000-0003-2791-6070}
\affiliation{%
  \institution{University of Twente}
  \city{Enschede} 
  \country{The Netherlands} 
}
\email{r.h.bemthuis@utwente.nl}

%The default list of authors is too long for headers}
\renewcommand{\shortauthors}{C. Pagels et al.}

\begin{abstract}
Enterprise Architecture Debt (EA Debt) arises from suboptimal design decisions and misaligned components that can degrade an organization's IT landscape over time. Early indicators, Enterprise Architecture Smells (EA Smells), are currently mainly detected manually or only from structured artifacts, leaving much unstructured documentation under-analyzed. This study proposes an approach using a large language model (LLM) to identify and quantify EA Debt in unstructured architectural documentation. Following a design science research approach, we design and evaluate an LLM-based prototype for automated EA Smell detection. The artifact ingests unstructured documents (e.g., process descriptions, strategy papers), applies fine-tuned detection models, and outputs identified smells. We evaluate the prototype through a case study using synthetic yet realistic business documents, benchmarking against a custom GPT-based model. Results show that LLMs can detect multiple predefined EA Smells in unstructured text, with the benchmark model achieving higher precision and processing speed, and the fine-tuned on-premise model offering data protection advantages. The findings highlight opportunities for integrating LLM-based smell detection into EA governance practice. 
\end{abstract}




\maketitle

\begin{center}
\footnotesize
This is the author's version of the work. The definitive Version of Record
appeared in \emph{Proceedings of the 41st ACM/SIGAPP Symposium on Applied
Computing (SAC '26)}. 
\end{center}

\section{Introduction}
Enterprise Architecture (EA) guides the design, evolution, and governance of an organization’s IT systems. Over time, however, architectures accrue inefficiencies, known as Enterprise Architecture Debt (EA Debt) that impair performance, scalability, and maintainability~\cite{hacks_towards_2019}. Similar to technical debt in software engineering, EA Debt reflects suboptimal design choices, outdated technologies, and misaligned components. These issues often manifest as Enterprise Architecture Smells (EA Smells), recognizable symptoms of potential debt accumulation. Left unaddressed, such debt can increase costs and reduce agility, thereby diminishing competitive advantage~\cite{hacks_towards_2019}. 

Identifying and analyzing EA Debt has been manual and resource-intensive so far. Common practices rely on workshops, interviews, and expert assessments to identify smells~\cite {jung_revealing_2021,daoudi_discovering_2023}. Although informative, these activities do not scale well (e.g., time-consuming, costly, and reliant on skilled architects)~\cite {jung_revealing_2021}. Moreover, the increasing complexity of modern IT systems, coupled with the sheer volume of architectural artifacts (e.g., documentation, codebases, and diagrams), has rendered manual analysis increasingly impractical~\cite{wang_viewpoints-based_2022}. As a result, there is a need for automated support to efficiently identify, analyze, and quantify EA Debt. 

Recent work has automated parts of this task by analyzing structured artifacts, such as models and knowledge graphs~\cite{serral_using_2021, alexanian_implementation_nodate}. Yet, many smells are semantic in nature (e.g., a temporary solution or ambiguous ownership in neutral comments) and do not manifest cleanly in structured data~\cite{alexanian_implementation_nodate}. Much of an organization’s architectural knowledge resides in unstructured sources, including narrative documents, policies, meeting notes, codebases, and architecture diagrams. These artifacts often contain key insights into design decisions, legacy system constraints, and the evolution of an organization’s IT landscape. Yet, these sources remain largely underused in current methods, limiting early detection of architectural issues. This significant gap in automated EA Debt detection may prevent organizations from proactively addressing inefficiencies, ultimately leading to increased costs and reduced agility. 

This work investigates whether Large Language Models (LLMs) can help bridge this gap by analyzing unstructured data to detect and quantify EA Debt. LLMs are effective at extracting and summarizing information from free text and may surface patterns that signal EA Smells~\cite{https://doi.org/10.1002/path.6232}. We adopt a Design Science Research (DSR) approach to build and evaluate a prototype that processes enterprise documents and identifies potential smells. By doing so, we provide a means to automate the identification of EA Smells. LLMs can provide enterprise architects with timely, data-driven insights into potential root causes of architectural degradation, enabling more proactive and cost-effective maintenance. 

The objective of this study is therefore to design, implement, and evaluate an LLM-based approach for detecting EA Smells in unstructured architectural documentation that relate to the business layer of EA. The aim is to assess the technical feasibility, effectiveness, and efficiency of such an approach in supporting EA Debt analysis. Given the small on-premise model and constrained hardware, we do not expect our artifact to outperform proprietary state-of-the-art LLMs. Instead, we examine whether a resource-constrained local model can still provide useful smell detection under strict data-protection and hardware limits, and how its behavior compares to a GPT-based baseline. 

This work makes the following contributions: (i) a problem analysis and set of functional and non-functional requirements for LLM-based detection of EA Smells in unstructured enterprise documentation; (ii) the design and implementation of an artifact that applies a fine-tuned LLM to identify candidate EA Smells; and (iii) an empirical evaluation using synthetic yet realistic business layer documents, including a comparative analysis with a proprietary LLM baseline, reporting on accuracy, processing time, scalability, and adaptability. 

Guided by a DSR process~\cite{johannesson_introduction_2014,venable_choosing_2017}, the paper is structured as follows. Section~\ref{section:background} reviews background and related work and, through a focused literature scan, explicates the problem of inefficient EA Debt detection in unstructured documentation. Section~\ref{subsec:requirements} outlines the requirements for an LLM-based EA Debt analysis tool by synthesizing functional and non-functional needs from related work. Section~\ref{subsec:design} designs and develops an on-premise, fine-tuned LLM prototype that operationalizes these requirements for 12 business layer EA Smells. Section~\ref{section:evaluation} demonstrates and evaluates the artifact on a synthetic case company using 30 business documents and benchmarks its performance against a custom Generative Pre-trained Transformer (GPT) baseline. Section~\ref{section:discussion} discusses implications, typical error modes, and threats to validity, and Section~\ref{section:conclusion} concludes. 

\section{Background and Related Work}
\label{section:background}

\subsection{Enterprise Architecture Debt and Enterprise Architecture Smells}
EA Debt extends the notion of technical debt to the context of enterprise architecture~\cite{hacks_towards_2019}. It captures the cumulative impact of suboptimal design choices, outdated technologies, and misaligned components that hinder maintenance, scalability, and adaptation of IT landscapes~\cite{hacks_towards_2019}. One formal definition describes EA Debt as: "the deviation of the currently present state of an enterprise from a hypothetical ideal state"~\cite[p. 201]{sales_towards_2024}. Such a deviation can arise from deliberate trade-offs under time or budget constraints, or from shifts in strategic priorities that render once-sound decisions less suitable over time~\cite{daoudi_discovering_2023}. EA Debt is therefore not inherently harmful as it can be a pragmatic instrument when managed explicitly~\cite{klinger_enterprise_2011}. Effective EA Debt management requires visibility into where debt exists and which items merit attention. 

EA Smells are symptoms that may indicate the EA Debt. Analogous to code smells, they serve as signs (or heuristic metric) for potential architectural issues in design or governance of enterprise systems~\cite {towards_a_catalog_2020}. Salentin and Hacks introduced a catalog of 45 EA Smells~\cite{towards_a_catalog_2020}, which provides a common vocabulary for detection and discussion. The catalog was extended to 63 smells by Lehmann et al.~\cite{lehmann_towards_2020} and Tieu and Hacks~\cite{tieu_determining_2021} by further adding business process anti-patterns and software architecture smells. Complementary to such catalogs, Jung and Hacks~\cite{jung2025taxonomy} developed taxonomies that systematically characterize and assess EA Debts, enabling a more structured evaluation of their context and impact. Similarly, Hacks and Slupczynski~\cite{hacks2025advancing} advanced the field by grounding EA Debt analysis in Work System Theory (WST), providing a theoretical lens on how debts arise and evolve across organizational systems. 

Existing automated approaches focus on structured EA artifacts and struggle with semantically rich business layer text~\cite{sales_exploring_2023,alexanian_implementation_nodate}. This work addresses that gap by investigating LLM-based detection of EA Smells directly from unstructured documentation. The study focuses on a scoped subset of business layer smells with high textual manifestation (cf. Section~\ref{section:artifact}), and evaluates both a resource-constrained, on-premise setup and a cloud baseline under data protection constraints. We assess feasibility, accuracy, and variance, and clarify where LLMs add value beyond structured-only analysis.

\subsection{Automated EA Analysis from Structured IT Artifacts}
Automation efforts in EA analysis have largely targeted structured artifacts. Prior work has shown detection of EA Smells or quality issues from enterprise models (i.e., ArchiMate) and knowledge graphs, leveraging formal structure and explicit relations~\cite{sales_exploring_2023,alexanian_implementation_nodate}. For example, Chis et al.~\cite{chis2024informing} show how WST concepts can be represented in a continuously updated Resource Description Framework (RDF) knowledge graph, creating a living digital twin of enterprise work systems that allows architects to query cross-layer dependencies and automatically detect and monitor EA Debts as the landscape evolves. Similarly, Flórez et al.~\cite{florez2016catalog} survey and re-implement over 50 automated analyses for enterprise models in an extended ArchiMate metamodel. Building on this line of work, Flórez et al.~\cite{Florez2014} propose an extensible, plug-in-based approach where ArchiMate models act as a living laboratory for automated analyses that detect, quantify, and prioritize flaws. 

These methods are powerful when models are complete and current. However, many business decisions and rationales are documented only in natural language. As a result, approaches limited to structured representations can miss signals embedded in, e.g., policies, process narratives, meeting notes, and strategy documents. This gap motivates methods that can analyze unstructured text alongside structured sources.

\subsection{LLMs for Unstructured Analysis}
LLMs have shown strong performance in extracting structured information from free text across domains, including healthcare~\cite{10.1371/journal.pone.0314136}, chemistry~\cite{D4CS00913D}, agriculture~\cite{peng_embedding-based_2023}, and social work~\cite{doi:10.1177/10497315241280686}. These results suggest that LLMs can surface latent patterns and entities in narrative documents relevant to EA analysis. 

Training LLMs from scratch is rarely feasible due to computational cost~\cite{10596333}. Instead, models are adapted via fine-tuning or prompting. Parameter-efficient fine-tuning (PEFT) updates only a small set of additional parameters, reducing resource needs while retaining base-model knowledge~\cite{liu2022fewshotparameterefficientfinetuningbetter}. Low-Rank Adaptation (LoRA) is a common PEFT technique that inserts trainable low-rank matrices into selected layers and has been shown to be effective in practice~\cite{hu2021loralowrankadaptationlarge}. Few-shot learning complements fine-tuning by guiding models at inference time with a handful of task-specific examples, which is useful when labeled data or hardware is limited~\cite{refId0}. Tooling ecosystems such as PyTorch and Hugging Face Transformers support these strategies and enable reproducible pipelines~\cite{10.1145/3710944}. 

Applying LLMs in enterprise contexts introduces risks. Data protection and confidentiality can preclude the use of external APIs, pushing toward on-premise deployment and smaller models~\cite{huang2025premises,paloniemi2025porting}. In addition, generation is stochastic, and outputs may vary across runs or input batching, which may complicate evaluation and governance~\cite{wang2025embedding}. Careful prompt and pipeline design~\cite{rodriguez2023prompts,chen2025unleashing}, explicit measurement of variance, and traceability to source text are therefore key when LLMs inform enterprise architecture decisions. 

\section{LLM-based EA Smell Detection Artifact}
\label{section:artifact}
\subsection{Problem Scope and Requirements}\label{subsec:requirements}
We develop a prototype that uses an LLM to identify, analyze, and quantify EA Debt from unstructured documents. Typical inputs are process descriptions, business rules, policy documents, and work instructions in \texttt{.docx} or \texttt{.pdf} formats. The prototype classifies EA Smells, and summarizes findings for potential EA Debt. Functional and non-functional requirements are summarized in Table~\ref{tab:functional_requirements} and Table~\ref{tab:nonfunctional_requirements}. Given the exploratory nature of this study, the non-functional requirements are kept broad to emphasize feasibility rather than precise performance thresholds. 

\begin{table}[ht]
  \centering
  \caption{Functional Requirements}
  \label{tab:functional_requirements}
  \begin{tabular}{p{0.35\linewidth} p{0.6\linewidth}}
    \toprule
    Requirement & Description \\
    \midrule
    Interpreting unstructured data sources & The artifact should be able to process Word (.docx) and PDF (.pdf) files. \\
    Detecting EA Smells & The artifact should recognize common architectural inefficiencies. \\
    Providing recommendations & The artifact should generate an output that includes the findings and recommendations to guide enterprise architects in addressing the identified issues. \\
    Data protection & Enterprise data should be protected. \\
    \bottomrule
  \end{tabular}
\end{table}

\begin{table}[ht]
  \centering
  \caption{Non-Functional Requirements}
  \label{tab:nonfunctional_requirements}
  \begin{tabular}{p{0.3\linewidth} p{0.65\linewidth}}
    \toprule
    Requirement & Description \\
    \midrule
    Technical limitations & The artifact needs to be able to run on a Windows 16 GB RAM environment with CPU only. \\
    Accuracy & The artifact should identify EA Smells correctly. \\
    Processing time & The artifact should have an adequate processing time when analyzing documents. \\
    Scalability & The artifact should be able to handle large datasets without performance degradation. \\
    Adaptability & The artifact should revise prior assessments based on newly provided, potentially corrective information. \\
    Usability & The artifact should be easy to use. \\
    Ethicality & The artifact needs to comply with ethical guidelines. \\
    \bottomrule
  \end{tabular}
\end{table}

\begin{figure}[ht]
\centering
\includegraphics[width=0.76\linewidth]{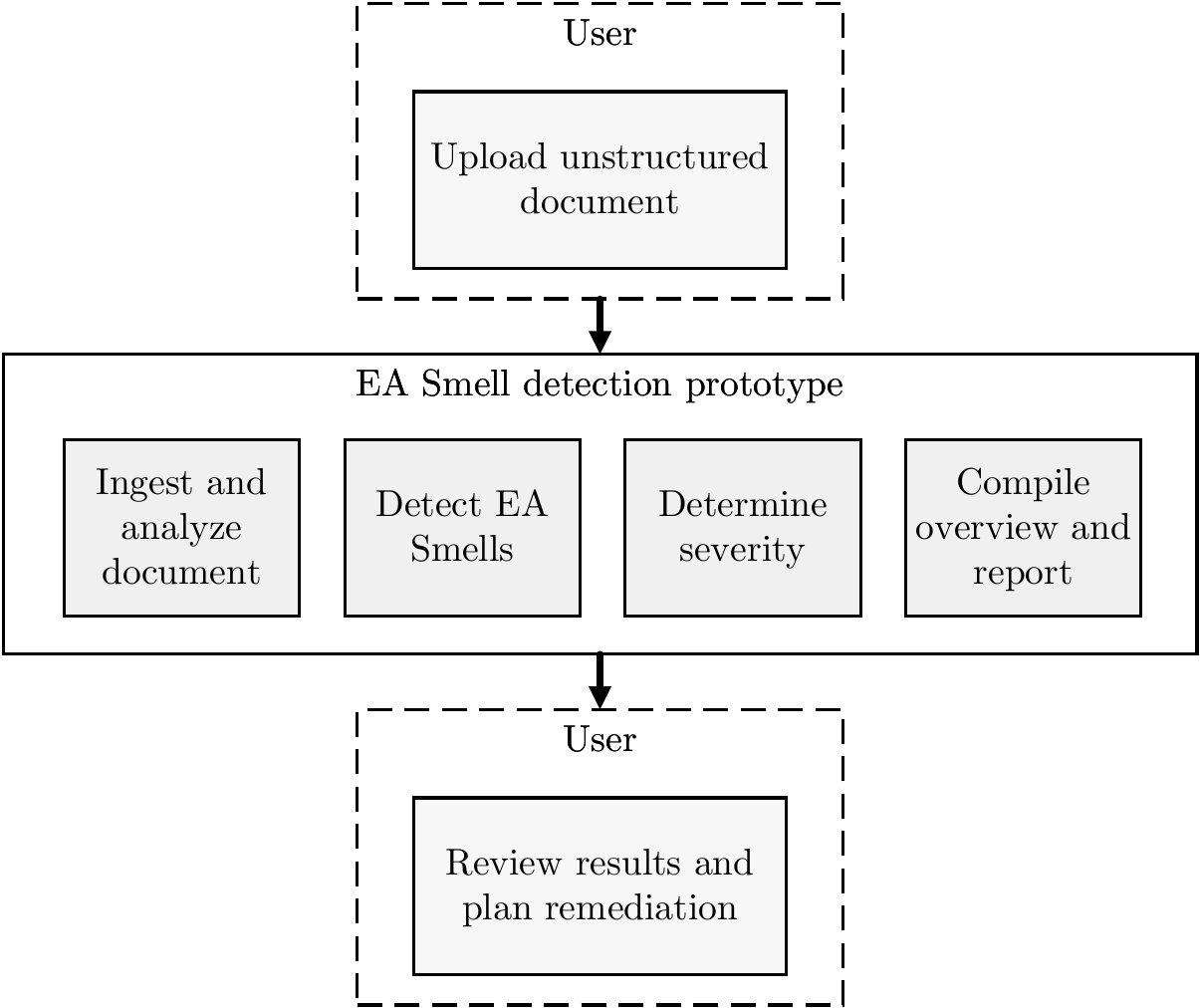}
\caption {Processing pipeline of the artifact.}
 \label{fig:artifactProcess}
\end{figure}

The pipeline (see Figure~\ref{fig:artifactProcess}) consists of: (i) ingestion and format normalization; (ii) preprocessing (tokenisation, document chunking with overlap to retain context); (iii) LLM-based smell detection and rationale extraction; and (iv) report generation with traceable snippets. To satisfy data-protection requirements, processing avoids external API calls and runs locally. 

\subsection{Design and Development}\label{subsec:design}
The artifact targets the business layer (processes, goals, roles, capabilities). Two considerations motivated this scope. First, business layer documentation is predominantly unstructured natural language and is less represented in structured repositories. Second, governance at this layer is often lighter than at the application and technology layers~\cite{info13010031}, where versioning and monitoring are stronger, latent debt is therefore more likely to go unnoticed. 

A comprehensive collection of 63 EA Smells has been defined and made available online. However, not all of them can be automatically detected using existing methods~\cite{alexanian_implementation_nodate}. Also, Jung et al. identified 12 further Smells in a workshop setting exhibiting potential for automated detection~\cite{jung_revealing_2021}. Combining these two sets results in 30 candidate Smells that were further evaluated for their feasibility in detection using LLMs. 

Only Smells with high feasibility were selected for the implementation scope, resulting in 12 Smells that the artifact will focus on, shown in Table~\ref{tab:smelloverview}. For example, the Smell "Contradiction in Input" presents clear textual signatures in policy and rule documents and exhibits low cross-document dependency. LLMs can identify logical inconsistencies, especially if the contradictory conditions are clearly described. A detailed breakdown of the feasibility assessment, including all evaluated Smells and their scores, is provided in an online repository~\footnote{\url{https://github.com/3492010/ea-debt-llm}}. 

\begin{table}[ht]
  \centering
  \caption{EA Smells Overview}
  \label{tab:smelloverview}
  \begin{tabular}{p{0.3\linewidth} p{0.65\linewidth}}
    \toprule
    EA Smell & Short Description \\
    \midrule
    Ambigious Viewpoint & Viewpoints are mixed or unclear. \\
    Big Bang & Architecture changed all at once. \\
    Business Process Forever & Processes kept unchanged despite need. \\
    Contradiction in Input & Inputs or rules contradict each other. \\
    Deficient Names & Names are unclear or misleading. \\
    Efficiency Goals not Visible & No measurable performance targets. \\
    Lack of Documentation & Key documentation missing. \\
    Language Deficit & Naming conventions are inconsistent. \\
    Project Goals not Achieved & Projects fail to meet objectives. \\
    Responsibilities not Defined & No clear ownership assigned. \\
    Shiny Nickel & New tech adopted without real need. \\
    Temporary Solution & Quick fixes become long-term. \\
    \bottomrule
  \end{tabular}
\end{table}

\subsubsection{Model Selection and Configuration}
\label{sec:artifact-model}
We explored open-source LLMs (e.g., LLaMA, Mistral, Gemma, Falcon) under the target hardware constraints. Models with $\geq$7B parameters were excluded due to CPU-only inference limits (memory and latency). The artifact must support instruction following and fine-tuning with common frameworks. Therefore, we considered models supported by PyTorch and Hugging~Face Transformers rather than GGUF-only builds. 

We selected \emph{LLaMA-3.2-3B-Instruct} for its instruction-following behavior, compatibility with parameter-efficient fine-tuning, and acceptable runtime on CPU within the 16~GB RAM budget~\cite{Llama3.2-3B-Instruct}. Cloud APIs (e.g., GPT-4) were not integrated into the artifact due to data-protection requirements. A proprietary comparator is used later only as a baseline (see Section~\ref{section:eval-baselines}). 

\subsection{Development Process} \label{developmentProcess}
This section outlines the key considerations and steps taken during the development of the artifact, the fine-tuning strategy, and dataset construction. The development was primarily conducted using Google Colab, leveraging its cloud-based resources to support prototyping and experimentation. While Colab provides limited access to GPU acceleration, the final artifact was designed to run in a CPU-only environment with 16 GB RAM, consistent with the non-functional requirements. 

The fine-tuning strategy consists of a combination of a few-shot learning and the LoRA approach, a specific PEFT technique. This approach is especially useful when working with lower technical requirements, as is the case for this prototype. By using LoRA, the base model's frozen weights are used while introducing small, trainable adapter layers. This method significantly reduces memory requirements while preserving the model's foundational knowledge. The public repository contains the full dataset and scripts.\footnote{\url{https://github.com/3492010/ea-debt-llm}}. 

We constructed a training dataset that covers the 12 target smells across eight business domains: Order-to-Cash, Procure-to-Pay, Record-to-Report, Hire-to-Retire, IT Management, Customer Service, Marketing, and Logistics. Each smell is represented by 80 examples (50\% positive, 50\% negative), yielding 960 instances in total. 

\subsubsection{Fine-Tuning Code}
A Python script was used in the Google Colab environment to fine-tune the LLaMA-3.2-3B base model for EA Smell identification. The final code was developed iteratively through a variety of public tutorials, extensive trial-and-error, and addressing a range of errors and limitations that emerged during experimentation. This process involved adapting the code to the technical constraints of the execution environment (memory and GPU availability), refining model parameters, and testing different configurations until a stable and functional setup was achieved. 

\subsubsection{Training Dynamics} \label{trainingloss}
We trained for four epochs per dataset. Loss curves (Figure~\ref{fig:training_loss}) show a monotonic decrease across 40 steps, with initial losses around 4.5--5.4 and final losses between 0.94 and 1.88. Before plotting, we corrected a logging scale issue caused by locale-specific decimal separators. 

\pgfplotsset{compat=1.18}

% Okabe–Ito colors (colorblind-safe)
\definecolor{oiOrange}{HTML}{E69F00}
\definecolor{oiSkyBlue}{HTML}{56B4E9}
\definecolor{oiBluishGreen}{HTML}{009E73}
\definecolor{oiYellow}{HTML}{F0E442}
\definecolor{oiBlue}{HTML}{0072B2}
\definecolor{oiVermillion}{HTML}{D55E00}
\definecolor{oiReddishPurple}{HTML}{CC79A7}
\definecolor{oiGrey}{HTML}{999999}

% Consistent, print-safe styling for all plots
\pgfplotsset{
  every axis plot/.append style={
    semithick,
    mark size=1.6pt,
    mark repeat=4,      % show a marker every 4th point
    mark phase=2,
    mark options={solid,draw=black}, % black outline for visibility in print
  },
  grid style={line width=.2pt, draw=gray!40},
  major grid style={line width=.3pt, draw=gray!55},
}

% 12 distinctive series: 8 unique colors (solid) + 4 with dashed variants
\pgfplotscreateplotcyclelist{oiCycle}{
  {oiBlue,           mark=*,           solid},
  {oiOrange,         mark=square*,     solid},
  {oiBluishGreen,    mark=triangle*,   solid},
  {oiVermillion,     mark=diamond*,    solid},
  {oiReddishPurple,  mark=star,        solid},
  {oiSkyBlue,        mark=o,           solid},
  {oiGrey,           mark=+,           solid},
  {black,            mark=x,           solid},
  {oiBlue,           mark=*,           dashed},
  {oiOrange,         mark=square*,     dashed},
  {oiBluishGreen,    mark=triangle*,   densely dashed},
  {oiVermillion,     mark=diamond*,    dotted},
}

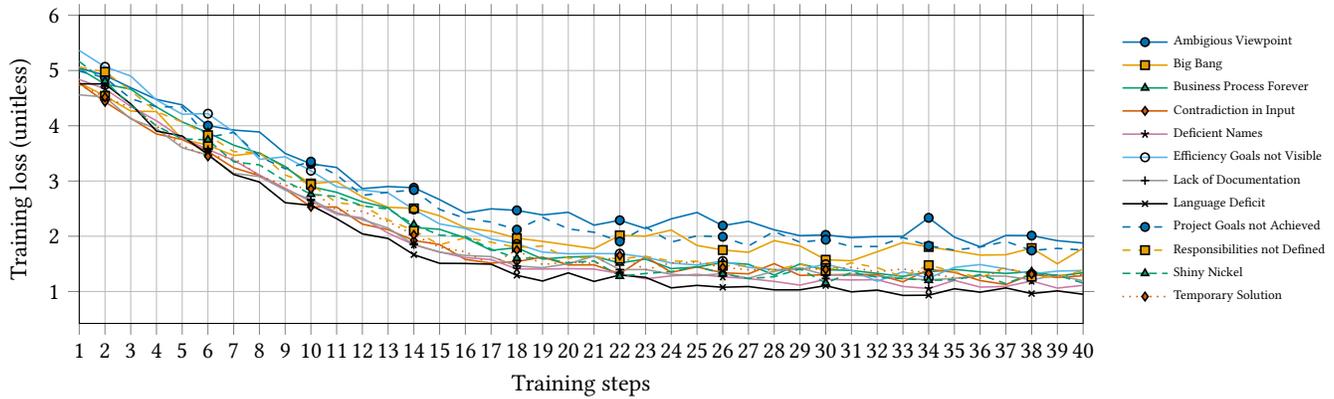
\begin{figure*}[ht!]
    \centering
\begin{tikzpicture}
\pgfplotsset{}
\begin{axis}[
    width=0.75\textwidth, % Adjust the width of the plot to take 75% of the total width
    height=4.1cm,
    scale only axis,
    tick pos=both,
    tick align=outside,
    xtick={1,2,...,40},
    ytick={1,2,...,6},
    xlabel={Training steps},
    ylabel={Training loss (unitless)},
    legend style={at={(1.03,0.5)}, anchor=west, legend columns=1, draw=none, font=\tiny},
    legend cell align=left,
    grid=both,
    enlarge x limits=false,
    xmin=1,
    xmax=40,
    ymax=6.0,
    cycle list name=oiCycle,
]
% Plot for Exp. 1
\addplot+ table [x=step, y=loss] {
step    loss
1	5.0437
2	4.9141
3	4.6882
4	4.4775
5	4.3777
6	4.0031
7	3.922
8	3.8878
9	3.4986
10	3.311
11	3.2449
12	2.8622
13	2.8997
14	2.8782
15	2.6632
16	2.421
17	2.4961
18	2.4683
19	2.3858
20	2.4336
21	2.2007
22	2.2892
23	2.1402
24	2.3146
25	2.4304
26	2.1919
27	2.27
28	2.1159
29	2.011
30	2.0245
31	1.9754
32	1.9934
33	1.9981
34	2.3338
35	1.9835
36	1.8094
37	2.0149
38	2.0118
39	1.9199
40	1.8774
};
\addlegendentry{Ambigious Viewpoint};

% Plot for Exp. 2
\addplot+ table [x=step, y=loss] {
step    loss
1	4.777
2	4.541
3	4.2657
4	4.257
5	3.7573
6	3.6479
7	3.461
8	3.5138
9	3.2409
10	2.9539
11	2.9884
12	2.7102
13	2.5246
14	2.5042
15	2.3696
16	2.1549
17	2.0919
18	1.9656
19	1.9072
20	1.8421
21	1.776
22	2.0158
23	2.0029
24	2.1124
25	1.8336
26	1.7512
27	1.7098
28	1.9209
29	1.8251
30	1.5749
31	1.5559
32	1.7191
33	1.886
34	1.8065
35	1.7366
36	1.659
37	1.6648
38	1.7856
39	1.5016
40	1.7834
};
\addlegendentry{Big Bang};

% Exp. 3
\addplot+ table [x=step, y=loss] {
step    loss
1	5.0345
2	4.7543
3	4.6621
4	4.341
5	4.0694
6	3.8753
7	3.6483
8	3.5015
9	3.2629
10	2.8984
11	2.795
12	2.6231
13	2.5135
14	2.1577
15	2.1265
16	1.9712
17	1.7446
18	1.793
19	1.5867
20	1.6231
21	1.638
22	1.5163
23	1.5868
24	1.4146
25	1.4418
26	1.5237
27	1.4925
28	1.2958
29	1.4995
30	1.3946
31	1.3809
32	1.3282
33	1.2727
34	1.3549
35	1.3995
36	1.3559
37	1.3273
38	1.3585
39	1.2811
40	1.3384
};
\addlegendentry{Business Process Forever};

% Exp. 4
\addplot+ table [x=step, y=loss] {
step    loss
1	4.7704
2	4.4333
3	4.1361
4	3.8519
5	3.7497
6	3.534
7	3.2388
8	3.0985
9	2.8526
10	2.5334
11	2.5281
12	2.2193
13	2.1197
14	1.9228
15	1.8503
16	1.5798
17	1.5151
18	1.5409
19	1.6219
20	1.4808
21	1.4819
22	1.3228
23	1.6292
24	1.3472
25	1.4442
26	1.338
27	1.3197
28	1.5046
29	1.2918
30	1.3018
31	1.3084
32	1.3061
33	1.1746
34	1.4018
35	1.3463
36	1.1978
37	1.1266
38	1.3083
39	1.2528
40	1.2866
};
\addlegendentry{Contradiction in Input};

% Exp. 5
\addplot+ table [x=step, y=loss] {
step    loss
1	4.8377
2	4.6663
3	4.3515
4	4.0899
5	3.775
6	3.571
7	3.3758
8	3.1087
9	2.8675
10	2.6316
11	2.4007
12	2.3305
13	2.0575
14	1.8471
15	1.7156
16	1.6124
17	1.579
18	1.4086
19	1.4057
20	1.4135
21	1.4064
22	1.3248
23	1.2293
24	1.2829
25	1.3104
26	1.2651
27	1.2336
28	1.1826
29	1.1154
30	1.2206
31	1.2091
32	1.2135
33	1.092
34	1.0535
35	1.2014
36	1.0797
37	1.0917
38	1.1892
39	1.0616
40	1.1123
};
\addlegendentry{Deficient Names};

% Exp. 6
\addplot+ table [x=step, y=loss] {
step    loss
1	5.3625
2	5.0672
3	4.8998
4	4.467
5	4.2091
6	4.2183
7	3.8885
8	3.3922
9	3.4382
10	3.1859
11	2.895
12	2.8365
13	2.7839
14	2.4833
15	2.2234
16	2.1421
17	1.9498
18	1.8607
19	1.7132
20	1.6835
21	1.6864
22	1.685
23	1.6222
24	1.5144
25	1.4801
26	1.5517
27	1.4378
28	1.3961
29	1.3971
30	1.4939
31	1.3833
32	1.1852
33	1.3576
34	1.2426
35	1.4374
36	1.4875
37	1.4088
38	1.3183
39	1.369
40	1.3805
};
\addlegendentry{Efficiency Goals not Visible};

% Exp. 7
\addplot+ table [x=step, y=loss] {
step    loss
1	4.5599
2	4.5193
3	4.1242
4	3.9476
5	3.6016
6	3.4687
7	3.1383
8	3.0809
9	2.8354
10	2.664
11	2.4275
12	2.3015
13	2.1481
14	1.848
15	1.7112
16	1.6516
17	1.6314
18	1.4613
19	1.4305
20	1.502
21	1.6405
22	1.3952
23	1.3989
24	1.3123
25	1.2864
26	1.3232
27	1.2413
28	1.3669
29	1.4275
30	1.2986
31	1.2924
32	1.2794
33	1.2375
34	1.2104
35	1.2382
36	1.2817
37	1.2785
38	1.2122
39	1.2969
40	1.1925
};
\addlegendentry{Lack of Documentation};

% Exp. 8
\addplot+ table [x=step, y=loss] {
step    loss
1	4.7586
2	4.7577
3	4.3963
4	3.9038
5	3.8138
6	3.4785
7	3.118
8	2.981
9	2.6075
10	2.5601
11	2.3182
12	2.0432
13	1.9598
14	1.666
15	1.5105
16	1.5072
17	1.4947
18	1.2928
19	1.1895
20	1.3372
21	1.1823
22	1.2974
23	1.2563
24	1.0656
25	1.111
26	1.0764
27	1.0896
28	1.03
29	1.03
30	1.1063
31	0.9931
32	1.0261
33	0.9292
34	0.933
35	1.0495
36	0.9878
37	1.0648
38	0.9624
39	1.0115
40	0.9499
};
\addlegendentry{Language Deficit};

% Exp. 9
\addplot+ table [x=step, y=loss] {
step    loss
1	4.9895
2	4.8695
3	4.4914
4	4.3438
5	4.3399
6	3.7883
7	3.88
8	3.4439
9	3.2217
10	3.3503
11	3.1094
12	2.7395
13	2.7977
14	2.8359
15	2.485
16	2.3231
17	2.2562
18	2.1192
19	2.3415
20	2.1338
21	2.0723
22	1.9061
23	2.1689
24	1.8945
25	2.0075
26	1.991
27	1.8249
28	2.0861
29	1.8921
30	1.9378
31	1.8122
32	1.8153
33	1.9777
34	1.8282
35	1.7484
36	1.8048
37	1.9074
38	1.7422
39	1.7807
40	1.75
};
\addlegendentry{Project Goals not Achieved};

% Exp. 10
\addplot+ table [x=step, y=loss] {
step    loss
1	5.0622
2	4.9758
3	4.6352
4	4.2546
5	4.0689
6	3.8288
7	3.5323
8	3.5109
9	3.1092
10	2.9398
11	2.6096
12	2.5545
13	2.2954
14	2.1038
15	1.8447
16	1.9742
17	1.8929
18	1.7971
19	1.8227
20	1.6016
21	1.62
22	1.5936
23	1.637
24	1.5489
25	1.5473
26	1.4664
27	1.464
28	1.3846
29	1.494
30	1.3784
31	1.5191
32	1.4318
33	1.2642
34	1.4769
35	1.3575
36	1.2495
37	1.4415
38	1.2687
39	1.2604
40	1.38
};
\addlegendentry{Responsibilities not Defined};

% Exp. 11
\addplot+ table [x=step, y=loss] {
step    loss
1	5.1634
2	4.8072
3	4.3682
4	3.999
5	3.7623
6	3.7441
7	3.3478
8	3.2912
9	2.9977
10	2.7611
11	2.7194
12	2.5492
13	2.4924
14	2.212
15	2.0196
16	1.9977
17	1.7667
18	1.6006
19	1.6077
20	1.5173
21	1.5536
22	1.2763
23	1.3263
24	1.3594
25	1.452
26	1.3229
27	1.2337
28	1.2675
29	1.4087
30	1.1537
31	1.3446
32	1.2586
33	1.2329
34	1.2026
35	1.2102
36	1.3051
37	1.1419
38	1.3197
39	1.2956
40	1.1528
};
\addlegendentry{Shiny Nickel};

% Exp. 12
\addplot+ table [x=step, y=loss] {
step    loss
1	4.7401
2	4.5175
3	4.328
4	3.9962
5	3.6381
6	3.4458
7	3.4189
8	3.0914
9	2.9228
10	2.8562
11	2.4663
12	2.4526
13	2.265
14	2.0242
15	1.7995
16	1.7197
17	1.4912
18	1.7509
19	1.4886
20	1.5327
21	1.5054
22	1.6549
23	1.394
24	1.5193
25	1.5308
26	1.4354
27	1.3946
28	1.3914
29	1.4209
30	1.3999
31	1.4374
32	1.3555
33	1.4093
34	1.3262
35	1.249
36	1.239
37	1.4146
38	1.3133
39	1.2807
40	1.3136
};
\addlegendentry{Temporary Solution};

\end{axis}
\end{tikzpicture}
\caption{Training loss for the set of EA Smells.}
\label{fig:training_loss}
\end{figure*}

\subsubsection{Creation of Custom GPT for Benchmarking}
To evaluate the output quality of the developed prototype based on the LLaMA3.2-3B model, a custom GPT was created using OpenAI’s ChatGPT Premium user interface. This custom GPT was configured to identify the same set of twelve EA Smells as targeted by the prototype. The objective of this step was to establish a reference point based on a well-established proprietary LLM that has a high accuracy and large popularity~\cite{10500411}.

The setup of the custom GPT was conducted within the web interface provided by ChatGPT Premium, which allows users to define a custom GPT by uploading documents, providing detailed instructions, and configuring system behavior. The same annotated training data that had been prepared for the fine-tuning of the LLaMA model was ingested into the custom GPT, where it was contextually referenced during inference. Unlike the Llama-based prototype, the GPT model did not undergo parameter updates; instead, it relied on in-context learning via uploaded files. This included labeled examples for each of the 12 EA Smells in scope. The ingestion was performed through direct upload within the GPT builder interface. Additionally, the system prompt used in the LLaMA-based prototype was reused in the custom GPT setup. This ensured that both models operate under the same high-level reasoning constraints and task framing, thereby enabling a more controlled and meaningful comparison of results.

After ingesting the training data and prompt, the custom GPT confirmed that it had successfully integrated the provided information and could perform the task as expected. While no training loss or performance metrics were available through the interface, interactive responses indicated that the system could recognize and classify the EA Smells when prompted with representative inputs. This custom GPT was then used in parallel with the LLaMA-based prototype to analyze the same documents from the case study.

\section{Evaluation}
\label{section:evaluation}
This section discusses a case-based evaluation of the artifact. The fictional firm \emph{NextTech} is a smart manufacturing headquartered in Sweden. It employs 5{,}500 staff, operates B2B across the EU, and develops IT systems. Its enterprise architects manage large volumes of business layer documentation, making manual review impractical. The study, therefore, examines whether automated detection can surface EA Smells from such unstructured artifacts. 

Before the experiments, we expected the custom GPT baseline (Model~B) to surpass the on-premise LLaMA model (Model~A) in precision and overall accuracy, while Model~A would act as a more sensitive triage tool with higher recall and more false positives. Our evaluation therefore focuses on characterizing these trade-offs under realistic deployment constraints rather than outperforming the GPT baseline. 

We compare two systems, an LLaMA-based prototype (Model~A) and a custom GPT configuration (Model~B). The dataset consists of 30 synthetic but realistic business documents created using DeepSeek AI, each containing embedded ground truth annotations. These annotations correspond to manually assigned smell labels. We evaluate the models under three protocols: (i) single-document processing (Run~1); (ii) three batches of ten documents (Run~2); and (iii) one batch of 30 documents (Run~3). Raw outputs and metadata are available in the public repository. 

\subsection{Design, Ground Truth, and Validity Notes}
Detected smells are verified against the embedded ground truth. We also record qualitative inconsistencies and error patterns. The dataset is synthetic and was annotated and evaluated by the same researcher. This raises risks of bias and overfitting to implicit assumptions. We mitigate this by reporting detailed failure modes and by separating quantitative from qualitative findings. 

Model outputs vary across sessions. Restarting a session and reprocessing the same input yields slightly different results, consistent with stochastic generation in LLMs. 

\subsection{Compared Systems and Protocol}
\label{section:eval-baselines}
\textbf{Model~A} is the fine-tuned LLaMA-based prototype evaluated on CPU-only hardware. \textbf{Model~B} is a custom GPT configured for the same twelve smells and used as a benchmark. The case study data set was run three times to
assess performance under increasing complexity conditions. For Run~1 we process each document independently. For Run~2 we process three batches of ten documents. For Run~3 we process a single batch of all 30 documents and can evaluate whether Model~A fulfills the scalability requirements. Model~B cannot accept more than ten documents per batch and is therefore excluded from Run~3. 

\subsection{Metrics}
We report accuracy, precision, recall, F1, and false positive rate (FPR). We also capture processing time, batch scalability, adaptability (ability to revise outputs given new context), usability (heuristic self-assessment), the specificity of recommendations, and error categories.

\subsection{Results}
\paragraph{Model~A, Run~1 (single documents).}
Model~A shows high sensitivity, flagging on average 10.5 smells per document; 23/30 documents contain 11–12 flagged items. However, the number of findings exceeds the prompt’s instruction to report only highly evident cases. Two patterns recur: (i) plausible but unintended smells are reported where none were embedded; (ii) planted smells are occasionally missed, with generic rationales in their place. Misclassifications occur. In Scenario~1, an O2C description is criticized for lacking project objectives despite not being a project document. \emph{Temporary Solution} and \emph{Big Bang} are often conflated. Missing documentation is sometimes labeled as \emph{Contradiction in Input} (e.g., Scenario~13). Output formatting is inconsistent: some results include recommendations, others do not; duplicate smells and duplicated recommendations appear (e.g., Scenarios~4, 15, 21). 

\paragraph{Model~B, Run~1 (single documents).}
Model~B correctly classifies 14/30 documents: ten without embedded smells and four with embedded smells (\emph{Lack of Documentation}, \emph{Shiny Nickel}, and \emph{Project Goals Not Achieved} twice). Sixteen documents are misclassified, mainly where two or three smells are embedded. The system also proposes additional smells in 15 documents, most often \emph{Shiny Nickel} and \emph{Ambiguous Viewpoint}, followed by \emph{Responsibilities Not Defined} and \emph{Efficiency Goals Not Visible}. All eight strategy documents are flagged for \emph{Shiny Nickel} and \emph{Ambiguous Viewpoint}, although only three embed these smells. Fourteen of these 15 additional flags are judged helpful, one is partially correct (issue detected but mapped to the wrong smell). 

\paragraph{Model~A, Run~2 (3 times of 10 documents).}
Batching may reduce verbosity and slightly lower the number of detected smells per document, but outputs become more generic and omit specifics. The misidentification of \emph{Shiny Nickel} and \emph{Temporary Solution} persists. Repetition within a document remains (e.g., Scenario~23). Some severities shift relative to run Run~1 (e.g., Scenario~19). Scenario~10 lists a non-existent smell (“Inefficient Goals Not Visible”). No cross-document summary is provided and results are listed sequentially. Scenario~28 echoes parts of the prompt instead of the results. Overall, batching yields fewer words but no tangible gain in precision. 

\paragraph{Model~B, Run~2 (3 times 10 documents).}
Accuracy drops to 11 correct classifications, nine of which contain no embedded smells. \emph{Lack of Documentation} and \emph{Responsibilities Not Defined} are detected, but four fewer documents are correct than in Run~1. Some smells missed in Run~1 are found in Run~2 (e.g., Scenario~20), while others are lost (e.g., Scenario~1). Additional suggested smells increase in some cases (e.g., Scenario~11: from two to four), and they remain plausible on inspection. One failure case attributes a smell to a citation not present in the focal document (the cited sentence belongs to another document in the same batch, suggesting context leakage across batch items). 

\paragraph{Model~A, Run~3 (1 times 30 documents).}
Results mirror those of Run~2. Outputs are shorter and more general. Accuracy remains low: planted smells are often missed, and misclassifications persist. Increasing batch sizes does not improve precision. 

\paragraph{Model~B, Run~3 (1 batch of 30 documents).}
Not applicable: the interface limits uploads to ten documents per batch. 

\paragraph{Detection Example}
In scenario 27, the inserted smell \emph{Project Goals not Achieved}, signaled by "Pilot project in Order Automation showed 20\% less turnover than expected because of late deployment", was correctly identified by Model~B (True Positive), but missed by Model~A (False Negative). Instead, Model~A incorrectly flagged \emph{Lack of Documentation} for IT Management and Hire-to-Retire performance, despite similar information density and structure across domains (False Positive). Additional details are available in the public repository. 

\subsection{Aggregated Metrics}
Table~\ref{tab:metrics_all} summarizes derived metrics for Runs~1–2. Model~A shows consistent over-detection: 20 false positives in Run~1 and 17 in Run~2, yielding FPR values of 0.95 and 1.00, respectively. Accuracy falls from 0.15 to 0.08; Precision is low (0.26 and 0.19); F1 scores are 0.24 and 0.15. These results indicate a mismatch between the model’s learned decision boundary and the smell definitions used for ground truth. 

\begin{table*}[ht]
  \centering
  \caption{Evaluation Metrics for Model A and Model B Across Runs 1 and 2}
  \label{tab:metrics_all}
  \begin{tabular}{p{0.2\textwidth} p{0.15\textwidth} p{0.15\textwidth} p{0.15\textwidth} p{0.15\textwidth}}
    \toprule
    \textbf{Metric} & \textbf{Model A – Run 1} & \textbf{Model A – Run 2} & \textbf{Model B – Run 1} & \textbf{Model B – Run 2} \\
    \midrule
    Accuracy & 0.15 & 0.08 & 0.40 & 0.42 \\
    Recall & 0.22 & 0.13 & 0.22 & 0.25 \\
    False Positive Rate (FPR) & 0.95 & 1.00 & 0.09 & 0.09 \\
    Precision & 0.26 & 0.19 & 0.88 & 0.89 \\
    F1 Score & 0.24 & 0.15 & 0.35 & 0.39 \\
    \bottomrule
  \end{tabular}
\end{table*}

Model~B is conservative and stable. Precision is high (0.88 and 0.89) with a low FPR of 0.09 in both runs. Recall is modest (0.22 and 0.25), with a slight improvement in Run~2. F1 scores are 0.35 and 0.39. In short, Model~B sacrifices recall for precision and produces few false positives. 

\subsection{Processing Time and Scalability}
Table~\ref{tab:processingTime} reports end-to-end throughput. Model~B processes one document in $\sim$2~seconds and a batch of ten in $\sim$2~seconds. Model~A requires $\sim$120~seconds per document and $\sim$17~minutes for ten. Model~B cannot process more than ten documents at once. Model~A processes one batch of 30 documents in about 50~minutes. 

\begin{table}[ht]
  \centering
  \caption{Processing Time per Run}
  \label{tab:processingTime}
  \begin{tabular}{p{0.3\linewidth} p{0.3\linewidth} p{0.3\linewidth}}
    \toprule
    \textbf{Number of Documents} & \textbf{Llama Prototype (Model A)} & \textbf{Custom GPT (Model B)} \\
    \midrule
    1 & $\sim$120 seconds & $\sim$2 seconds \\
    10 per batch & $\sim$17 minutes & $\sim$2 seconds \\
    30 per batch & 50 minutes & Not possible \\
    \bottomrule
  \end{tabular}
\end{table}

\subsection{Adaptability}
Model~B occasionally revises earlier findings when follow-up documents add clarifying context (e.g., missing information is later supplied). This behavior is most visible in Run~1. It is less reliable in batch settings, where contextual boundaries are weaker. Model~A does not revise prior outputs when new documents are added. It also does not link related content across documents in the same batch (e.g., Scenarios~15 and the corresponding process description), indicating limited cross-document reasoning. 

\subsection{Usability}
Model~B’s conversational interface is responsive, supports flexible queries, and delivers concise answers quickly. Model~A’s Python interface requires manual I/O handling, provides limited interactive feedback, and is slower, which increases effort and cognitive load. These differences matter in iterative analysis settings. 

\subsection{Recommendation Quality}
Both models generate recommendations conditioned on the detected smells. Model~A tends to produce general actions (e.g., in Scenario~1 for the smell "Efficiency Goals not Defined": "Review the O2C process and identify clear efficiency targets and KPIs. Provide a clear plan for measuring and tracking these metrics"). Model~B often proposes more specific guidance for the same case: 

\begin{quote}
\itshape
Enterprise architects should introduce granular KPIs across all O2C stages to ensure full process visibility and performance management. Suggested additions include: 

\begin{itemize}
  \item Order Validation Time (e.g., target <2 hours from entry to approval)
  \item Billing Error Resolution Time (e.g., mean time to issue credit/refund)
  \item Dispute Resolution Accuracy (e.g., percentage of resolved disputes without recurrence within 30 days)\end{itemize}
\end{quote}

Both types of recommendations can assist enterprise architects by providing initial mitigation ideas. However, as the outputs are generated ad hoc and may vary across runs, they should be treated as supportive guidance rather than prescriptive solutions and require expert validation.

\subsection{Error Analysis}
We observe five recurring error types: (1) \textbf{Omission}: embedded smells are missed, especially in multi-smell documents; (2) \textbf{Misclassification}: a valid issue is mapped to the wrong smell; (3) \textbf{Batch context leakage}: text from one document is attributed to another within the same batch; (4) \textbf{False positives}: issues are flagged that are not justified by the document content; and (5) \textbf{Fabricated citations}: rationales cite sentences that do not appear in the referenced document. 

These errors point to gaps in smell disambiguation, prompt isolation, and traceability. They also indicate over-generalization in Model~A and context-boundary fragility in Model~B. 

\section{Discussion and Implications}
\label{section:discussion}
The research question guiding this study was: \textit{How can LLMs be used to identify EA Smells in unstructured architectural documentation?} The results indicate that LLMs have the potential to surface smell-related signals in business artifacts, such as process descriptions and strategic plans. The fine-tuned LLaMA-3 artifact provided end-to-end automation, but with modest accuracy and many false positives. By contrast, the custom GPT benchmark yielded substantially higher precision (approximately 0.88) and a lower false positive rate (approximately 0.09), alongside faster responses and better usability. Its recall and stability under batch input remained limited. Overall, effectiveness depends on data quality, prompt design, model choice, and evaluation protocol. 

\subsection{Interpretation of Model Behavior}
Both models identified plausible issues but differed in error modes. The LLaMA-based artifact over-detected smells and often produced generic explanations. This pattern is consistent with a mismatch between training examples and the semantics of real documentation and with prompts that encourage broad coverage. In one case, a standard Order-to-Cash description was flagged as \emph{Shiny Nickel}, suggesting limited contextual grounding. 

The custom GPT was more conservative. It missed some planted smells but produced a few false positives. It occasionally surfaced concerns not present in the ground truth. These cases can be useful early warnings, yet they may also reflect hallucinations or gaps in labeling. In batch settings, both models showed context bleed across documents, which reduced traceability and reproducibility. 

\subsection{Implications for EA Practice}
First, LLM outputs should be used as triage signals, not definitive diagnoses. Architects should review findings, confirm evidence in the source text, and decide follow-up actions. Second, privacy and deployment constraints matter: the custom GPT offers strong precision and speed but may not suit sensitive environments; on-premise models better address data control at the cost of accuracy and throughput. Third, workflow design is key. Processing documents individually, anchoring detections to quoted spans, and storing decision logs improve auditability. Finally, prompt hygiene and domain glossaries help reduce spurious flags and improve consistency of recommendations. 

\subsection{Limitations and Threats to Validity}
This study relied on synthetic corpora to protect confidentiality. Such data may not capture the nuance and noise of real documentation, limiting external validity. Additionally, we make the assumption that the documentation is up-to-date, which might be limited in reality. The training set covered 12 smells with 960 examples, which constrains generalization. Technical constraints (CPU-only, small models) restricted the range of architectures explored. Batch processing introduced context leakage across documents. Finally, accuracy assessments were manually verified by a single researcher. Hence, the absence of inter-rater reliability and confidence intervals limits the strength of the conclusion. These threats motivate the use of larger, more diverse datasets, multi-rater labeling, and variance-aware evaluation in future work.

\section{Conclusion}
\label{section:conclusion}
This study examined the use of LLMs to detect EA Smells in unstructured business documentation. The LLaMA-3 artifact demonstrated feasibility under strict deployment constraints but produced many false positives. From a pure detection-performance perspective, Model~A is therefore not yet suitable for production use, but it provides a baseline and exposes failure modes that inform the design of future on-premise EA analysis tools. The custom GPT benchmark achieved higher precision and lower false positive rates, with clear usability benefits, but showed modest recall and instability under batched inputs. The findings suggest that LLMs can support EA governance as decision aids when embedded in human-in-the-loop workflows and complemented with provenance and validation steps. This work extends prior research on EA Debt by focusing on unstructured artifacts and by comparing an on-premise model with a cloud-hosted benchmark on the same detection task. The results underscore the need for improved grounding, more robust labeling protocols, and hybrid architectures that integrate textual analysis with structured repositories. A future direction to improve performance could be to translate unstructured documents into Knowledge Graphs first and then analyze those instead.

Future studies should prioritize dataset curation and label quality, including multi-annotator protocols and adjudication to strengthen ground truth. Evaluation should report variance over multiple runs (with fixed seeds and prompt versions) and include ablations on batch size and few-shot context. Future research should expand and diversify training data, investigate hybrid pipelines that combine unstructured analysis with structured sources (e.g., ArchiMate models, knowledge graphs), and add retrieval-based grounding with span-level evidence to reduce context leakage and anchor claims. These steps may lower false positives, improve explanations, and enhance interpretability. Additional directions include testing larger models where feasible, supporting more file types (e.g., slide decks), and collaborating with industry partners to validate the approach on real documentation and assess its practical value. 

\printbibliography

\end{document}